\documentclass[a4paper,11pt]{article}
\usepackage{jinstpub} 
\usepackage{lineno}
\usepackage{subfig}
\usepackage{comment}
\usepackage{tikz-feynman} 

\newcommand{\mycomment}[1]{}{\bf }
\newcommand{\met}{$E_\text{T}^{miss}$}

\usepackage{float}

\notoctrue

\newcommand{\mbb}{M_{bb}}
\DeclareOldFontCommand{\rm}{\normalfont\rmfamily}{\mathrm}
\DeclareOldFontCommand{\sf}{\normalfont\sffamily}{\mathsf}
\DeclareOldFontCommand{\tt}{\normalfont\ttfamily}{\mathtt}
\DeclareOldFontCommand{\bf}{\normalfont\bfseries}{\mathbf}

\title{Enhancing Sensitivity for Di-Higgs Boson Searches Using Anomaly Detection and Supervised Machine Learning Techniques}

\author[a]{S.~Chekanov,}
\author[b]{W.~Islam,}
\author[a]{and N.~Luongo}

\affiliation[a]{HEP Division, Argonne National Laboratory, USA}
\affiliation[b]{Department of Physics, University of Wisconsin, Madison, WI, USA.}

\emailAdd{chekanov@anl.gov, wasikul.islam@cern.ch, nluongo@anl.gov}

\abstract{
This paper explores different strategies for enhancing sensitivity to new heavy resonances that decay into two or more Higgs bosons. This is achieved using two neural network architectures: an unsupervised autoencoder for anomaly detection and a supervised classifier. The autoencoder is trained on a small fraction of Standard Model (SM) Monte Carlo simulated events to calculate the loss distribution for input events, aiding in determining the extent to which events can be considered anomalous. The supervised classifier uses the same inputs but is trained on events simulated using both beyond Standard Model (BSM) and SM processes.
By applying selection cuts to the output scores, we compare the sensitivities of the two approaches.
}


\arxivnumber{2504.12418} 

\note{Preprint: HEP-ANL-195813 }

\begin{document}

\maketitle
\flushbottom

\clearpage


\section{Introduction}
\label{sec:intro}

In the Standard Model (SM), the Higgs boson is responsible for giving mass to other elementary particles through the Higgs mechanism. Studying di-Higgs production (i.e. the production of two Higgs bosons) can provide insights into the self-interactions of the Higgs field.  Many theories beyond the Standard Model (BSM), such as supersymmetry, composite Higgs models, extra dimensions, and others, predict deviations from the SM  in the Higgs sector. These could lead to enhanced rates for di-Higgs production, or produce enhancement in the invariant masses of the two Higgs bosons~\cite{Binoth:2006ym, PhysRevD.107.034014}.  Thus, it is important to study invariant masses of two Higgs bosons.  
To date, SM di-Higgs production has not been observed at the Large Hadron Collider (LHC). However, its observation is anticipated during the High-Luminosity phase of the LHC.

The most common approach for searching for new physics, whether in a general context or specifically within the di-Higgs sector, is to design an event classifier using neural networks or boosted decision trees.
Such machine learning (ML) tools are trained on events generated by Monte Carlo models, representing both BSM and SM processes. While SM event modeling is well established, the dependence of these algorithms on particular BSM assumptions constrains the generality of the resulting searches.

Reliance on BSM physics simulations can be avoided when creating neural networks that ``remember'' the primary kinematic characteristics of collisions events which are predominantly governed by SM processes. If such networks are trained
on a small fraction of well-understood data, or SM Monte Carlo and applied to data, the events which are recognized by such networks can be removed, leaving only those that cannot be fully reconstructed. In this case, no prior knowledge of BSM models is required, making this selection method more agnostic to potential new physics.  An overview of anomaly detection methods is provided in \cite{BELIS2024100091, Farina2018AE, Heimel2019AD, Mikuni2019VAE, Nachman2020Density, Guest2018Review, ATLAS:2023ixc, CMS2023AD}. ATLAS recently published the first LHC article utilizing anomaly detection to identify collision events that are most likely to contain BSM scenarios \cite{ATLAS:2023ixc}. This article considered five types of reconstructed objects (jets, $b$-jets, electrons, muons and photons) in its reconstruction of such events.

This article compares different strategies for enhancing sensitivity to new heavy resonances decaying to two or more Higgs bosons.
This was done using two neural network architectures: an unsupervised autoencoder for anomaly detection (AD) and a supervised classifier (SC). The autoencoder is trained on a small fraction of SM Monte Carlo events to calculate the loss distribution for input events, determining the degree to which the events can be considered anomalous. 
The supervised classifier uses similar inputs, but is trained on BSM and SM events. The performance of the SC is tested on events generated from a BSM model from which it has seen no training data. This is then compared with the performance of the AD model on those same BSM events, allowing for a comparison of the robustness of the two techniques.



\section{Monte Carlo Simulations}
\label{sec:simulation}

Even if the exact BSM process is well known, non-ML methods require significant effort to identify useful variables and design selection cuts, especially when the expected signal involves multiple final-state objects. ML methods, such as event classifiers, mitigate these challenges, but neural networks must be trained on Monte Carlo samples generated with the specific SM and BSM processes. 
The latter is affected by variations in the BSM modeling.  Anomaly detection, in contrast, does not require prior training on BSM models; instead, it seeks regions of phase space most sensitive to unusual activity that could (potentially) be populated by BSM signals. 
In some situations, this gives anomaly-detection approaches an edge. On the other hand, anomaly-detection methods may be less sensitive to certain BSM signatures, which are better studied using techniques that incorporate full information about the expected exotic processes.

For the comparison between event classification and anomaly detection, we have chosen  decays of Higgs-boson pairs that yield complex final states involving leptons, light-flavor jets, and jets associated with $b$-quarks.
Several BSM Monte Carlo event samples are used as outlined below:

\subsection{Signal models}
The current analysis uses two exotic processes with a heavy scalar $X$ decaying
into other scalars.  The first process is:
\begin{equation}
    X\rightarrow HH \rightarrow  VV b\bar{b}   \rightarrow l lb\bar{b} + A,
    \label{eq:hh}
\end{equation}
where $H$ denotes the SM Higgs boson. The Feynman diagram is shown in Fig.~\ref{fig:Xtohh_vs_XtoSh}(a). The last part of Eq.~\ref{eq:hh} indicates the decay mode, where one Higgs boson decays to vector bosons ($V$ = $Z, W$), while the second Higgs boson decays to two $b$-initiated jets (i.e. $b\bar{b}$). The two leptons ($ll$) in the final state
are expected to originate from either the $V=Z$ boson ($Z\rightarrow l^+l^-)$  or 
from semi-leptonic decays of two $V=W$ bosons. 
No requirements were set on the charge of the leptons during the analysis.
The final states contain neutrinos from decays of the VV bosons, which lead to missing transverse energy \met. 
The symbol $A$ in Eq.~\ref{eq:hh} indicates other decay products of the vector bosons, which could be \met from neutrinos, extra jets etc. While information on these decays is included in the ML algorithms used in this paper, it is not directly employed for signal reconstruction.
The selected Higgs decay channel, though less sensitive than other modes such as 
($H\rightarrow \gamma\gamma$), provides a complex final state that serves as an excellent testbed for developing and evaluating advanced ML techniques.

The event samples with the $X\rightarrow HH$ process were created at leading-order QCD with the {\sc Pythia8} Monte Carlo generator   (version  8.307) \cite{Sjostrand:2007gs} using $pp$ collisions at $\sqrt{s}=13.6$~TeV.
The generated masses for the heavy scalar $X$ were set to 0.5, 0.7, 1.0, 1.5, and 2~TeV.
The cross sections were set to the model double Higgs models as predicted by the  {\sc MadGraph5\_aMC@NLO}~\cite{Alwall:2014hca} at next-to-leading
order in QCD~\cite{Degrande:2015vpa}. A sample of 0.5 million events 
was created by allowing all Higgs decays. The generated events were preselected with at least two leptons at transverse momenta of $p_T>14$ GeV.

In this analysis, the ``signal'' region represents the invariant mass of two $b-$associated jets, which will be denoted by $\mbb$. This invariant mass should exhibit a sharp peak near the Higgs mass of 125~GeV.

The second signal sample was created from the model \cite{Chen:2022vac}:

\begin{equation}
    X\rightarrow SH \rightarrow  HHH   \rightarrow llb\bar{b} + A
    \label{eq:hhh}
\end{equation}
where a new scalar $S$ decays to two other SM Higgs bosons. The Feynman diagram is shown in Fig.~\ref{fig:Xtohh_vs_XtoSh}(b). This process was also generated with {\sc Pythia8} using $pp$ collisions at $\sqrt{s}=13.6$~TeV. 
This particular decay chain was recently studied in \cite{low2022higgsalignmentnovelcpviolating}.
As previously noted, the symbol  $A$ refers to the other decay products of $HHH$, such as  \met, extra jets etc, 
which are incorporated into the machine learning inputs.
The generated masses for the heavy scalar $X$ were set to the same values as for the process in Eq.~\ref{eq:hh}.
The mass of the scalar $S$ was set to 280~GeV, which is slightly above the mass threshold of the
$HH$ system. This selection of the mass provides the largest cross section for this channel.
Then the generated events were preselected with at least two leptons at $p_T>14$ GeV.

\begin{figure}[t]
\centering
 \subfloat[$X\rightarrow HH$]
 {
\centering
\begin{tikzpicture}
\begin{feynman}
  \vertex (g1) at (-2.0,  0.8) {\(g\)};
  \vertex (g2) at (-2.0, -0.8) {\(g\)};
  \vertex (v1) at (-0.6, 0);
  \vertex (v2) at (1.0, 0);
  \vertex (h1) at (3.0, 0.8) {\(H\)};
  \vertex (h2) at (3.0,-0.8) {\(H\)};

  \diagram*{
    (g1) -- [gluon] (v1) -- [gluon] (g2),
    (v1) -- [scalar, edge label=\(X\)] (v2),
    (v2) -- [scalar] (h1),
    (v2) -- [scalar] (h2)
  };
\end{feynman}
\end{tikzpicture}
}

 \subfloat[$X\rightarrow SH$] {
\begin{tikzpicture}
\begin{feynman}
  \vertex (g1) at (-2.0,  0.8) {\(g\)};
  \vertex (g2) at (-2.0, -0.8) {\(g\)};
  \vertex (v1) at (-0.6, 0);
  \vertex (v2) at (1.0, 0);
  \vertex (h1) at (2.8, 1.0) {\(H\)};
  \vertex (v3) at (2.4,-0.6);
  \vertex (h2) at (4.0, -0.1) {\(H\)};
  \vertex (h3) at (4.0, -1.1) {\(H\)};

  \diagram*{
    (g1) -- [gluon] (v1) -- [gluon] (g2),
    (v1) -- [scalar, edge label=\(X\)] (v2),
    (v2) -- [scalar] (h1),
    (v2) -- [scalar, edge label'=\(S\)] (v3),
    (v3) -- [scalar] (h2),
    (v3) -- [scalar] (h3)
  };
\end{feynman}
\end{tikzpicture}
}

\caption{Representative Feynman diagrams for: (a) the direct decay \(gg \!\to\! X\!\to\! H\,H\);  and (b) the cascade decay \(gg \!\to\! X\!\to\! S\,H \!\to\! H\,H\,H\).}
\label{fig:Xtohh_vs_XtoSh}
\end{figure}

\subsection{Background sample}

The main SM background for the $HH$ and $HS$ processes is events with two leptons from top-quark production. Double ($t\bar{t}$) and single ($t$) top production processes were created with {\sc Pythia8}. 
All top decays were enabled, but events were 
selected with at least two leptons at $p_T>14$~GeV.
The second largest background is $W/Z$+jet production.  It was also generated using {\sc Pythia8}.
Both SM processes were scaled to their corresponding cross sections according to the SM.

\subsection{Event processing for  ML inputs}

Following event generation with the requirement of two leptons with $p_T > 14$~GeV, all analysis objects were derived from truth-level particles.
The jets were constructed with the anti-$k_T$ algorithm \cite{Cacciari:2008gp} as implemented in the {\sc FastJet} package~\cite{Cacciari:2011ma} with a distance parameter of $R=0.4$, which is commonly used in the LHC experiments. By default, the minimum transverse energy of all jets was $20$~GeV in the pseudorapidity range of $|\eta|<2.5$. Leptons were required to be isolated using a cone of size $0.2$ in the azimuthal angle and pseudorapidity defined around the true direction of the lepton. All energies of particles inside this cone were summed. A lepton is considered isolated if it carries more than $90\%$ of the cone energy. The SM background processes require simulations of misidentification rates for muons and electrons (``fake rates''). We used a misidentification rate of 0.1\% for muons, and 1\% for electrons. They are implemented by assigning a probability of $10^{-3}$ ($10^{-2}$) for a jet to be identified as a muon (electron) using a random number generator. 

The signal region of both processes is based on the measured $bbll$ final state. 
As the final step, all events were selected by requiring two $b$-associated jets with $p_T>20$ GeV and two leptons above $15$~GeV.

\begin{figure}[H]  
  \begin{center}
    \subfloat[$X\rightarrow HH$]{\includegraphics[width=0.49\textwidth]{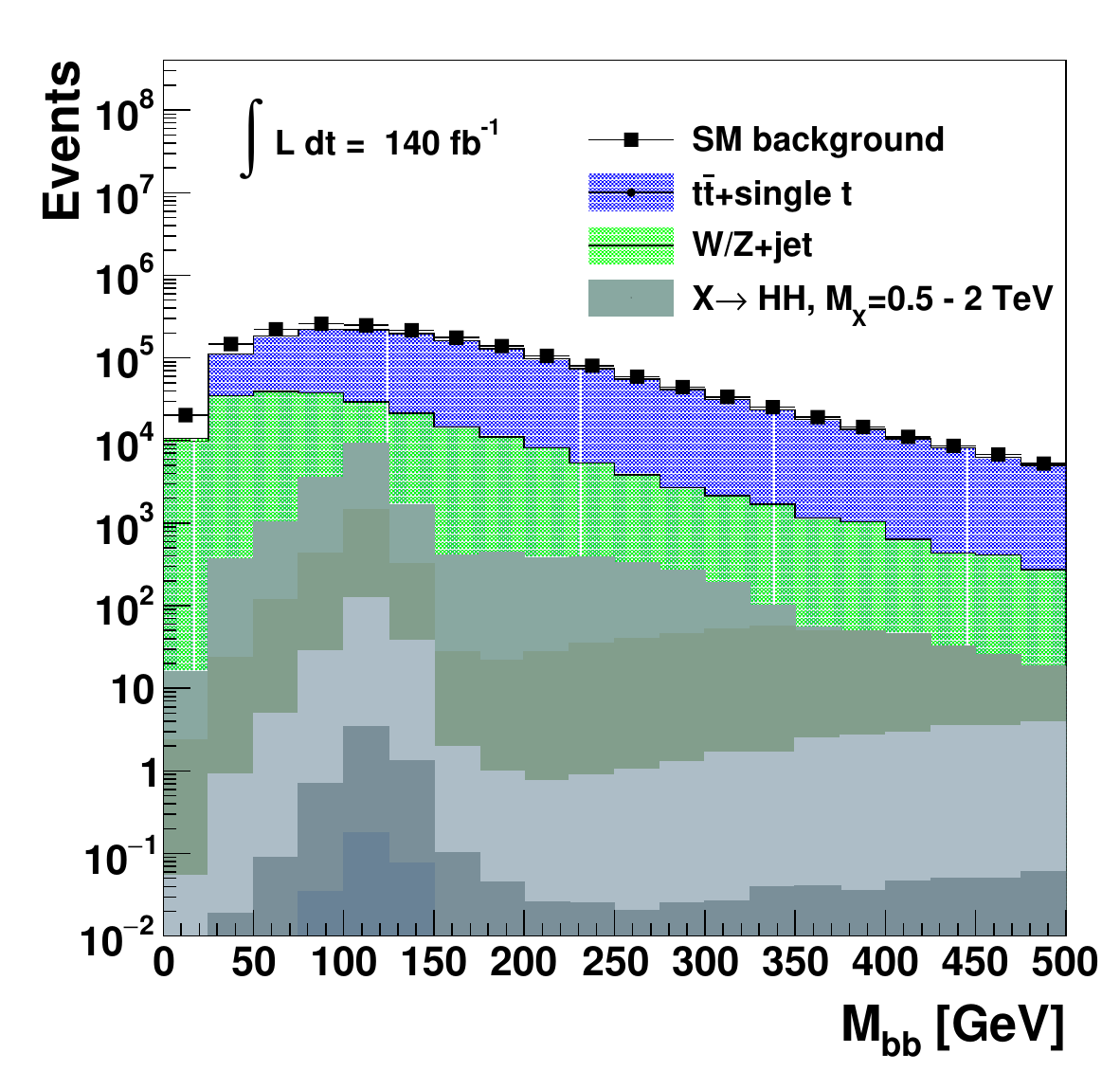}}
     \subfloat[$X\rightarrow SH$]{\includegraphics[width=0.49\textwidth]{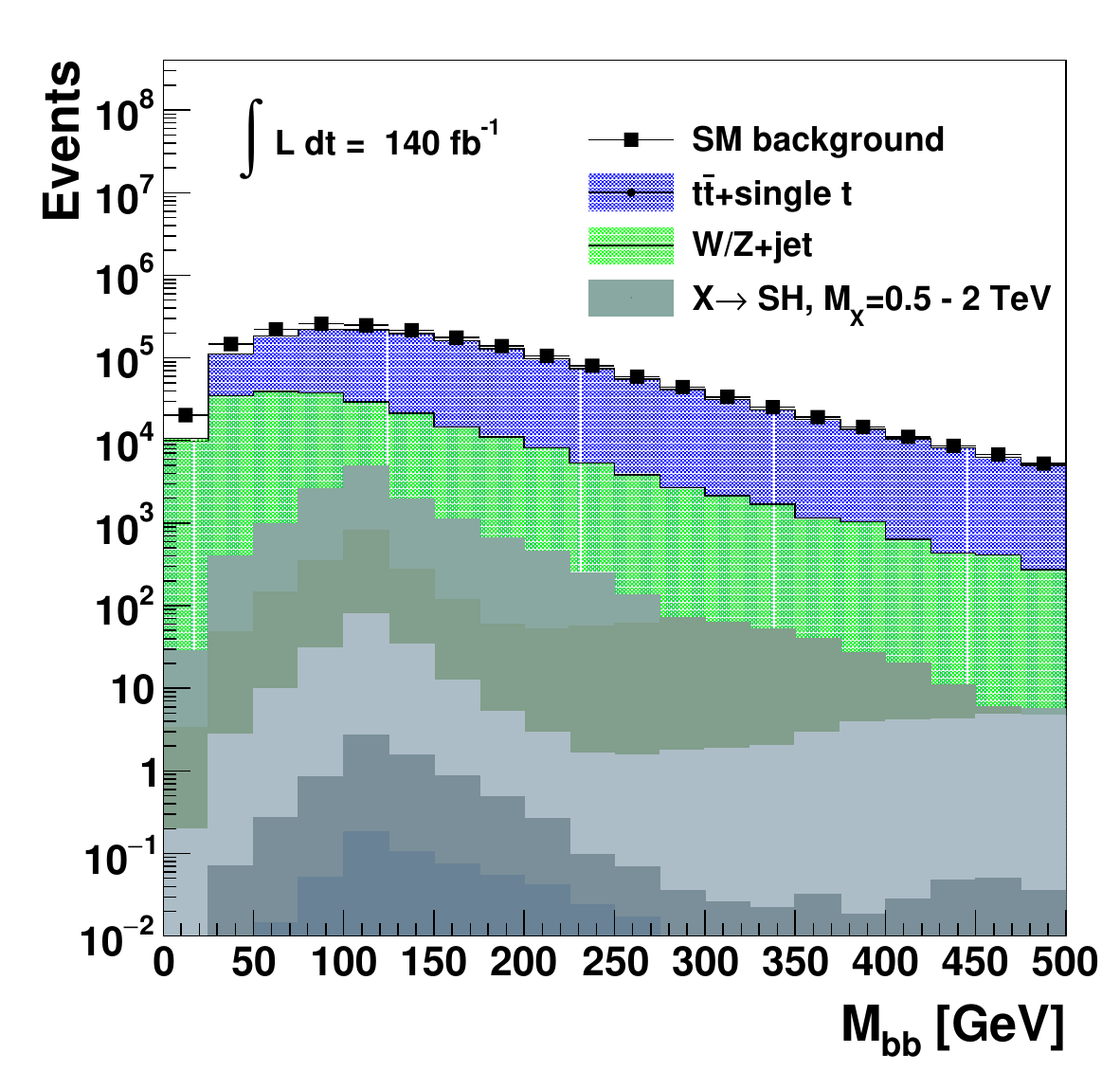}}
  \end{center}
\caption{The $\mbb$ invariant mass for $X\rightarrow HH$ and  $X\rightarrow SH$ processes before applying ML algorithms. The cross section for the latter process was set to that of $X\rightarrow HH$. Different shades of the gray color show the BSM signals with different masses of $X$, i.e. 0.5, 0.7, 1, 1.5 and 2 TeV. The larger the mass of $X$, the smaller area under the shaded histograms is expected.}
\label{fig:kin_mbb_before}
\end{figure}

Figure~\ref{fig:kin_mbb_before} shows the signal region with the invariant mass of two $b-$jets for the models $HH$ and $SH$, together with the expected SM backgrounds.  The predictions for the number of signal events were obtained for the 140~fb$^{-1}$ of an integrated luminosity.
The number of signal events for the model $SH$ was set to that of the model $HH$ in order to simplify the comparisons for the significance calculation, which will be discussed later. 
As expected, both BSM models show a peak near the Higgs mass of 125~GeV. The signal from the $SH$ model is somewhat broader, due to a larger combinatorial background arising from a mis-assignment of $b-$jets from the Higgs decays.  The signal events are difficult to observe because of the overwhelming contributions from the SM background near the 125~GeV region.

The number of $W/Z$+jet events in the signal region is roughly an order of magnitude smaller than the number from the top processes. Furthermore, it has been observed that $W/Z$+jet events are generally much easier to separate from signal events compared to top process events. For these reasons and for simplicity, the SC and AD models are trained using only top process events as background.

After all objects (jets, $b-$jets, leptons and photons) 
are available, a C++ program is used to transform kinematic features of each event
to the rapidity--mass matrix (RMM) which is provided as an input for ML~\cite{Chekanov:2018nuh}. 
These inputs provide a generic, model-agnostic representation of event kinematics and therefore do not require handcrafting process-specific variables, as is often necessary when relying on traditional high-level observables. The main advantage of the RMM is that all inputs are normalized to the range (0, 1) and exhibit monotonically decreasing (“falling”) distributions, which tend to be more stable \cite{Chekanov:2021pus} as machine-learning inputs than the distributions of raw object four-momenta, which typically feature a sharp peak at 0 with symmetric Lorentzian tails. Other input styles, such as object transverse momentum and pseudorapidity, are subsets of variables already included in the RMM.

The RMM is a square matrix that includes reconstructed final states of jets, $b$-jets, muons, electrons, photons, and \met, where \met is a single object, followed by $10$ ($b$-)jets and $5$ electrons, muons and photons each, in descending order of transverse energy for each particle type. By construction, all elements of the RMM are defined to be between 0 and 1, and most variables are Lorentz-invariant under boosts along the longitudinal axis. 
To reduce biases in the shapes of the jet\;\!+\;\!$Y$ invariant mass spectra, the nine invariant mass variables are excluded from the RMM. The resulting input dimension is $36^2 - 9 = 1287$.
The RMM matrix is then flattened into a one-dimensional input vector before being fed into the AE or SC.

The RMM was first proposed for backpropagation-based event classification algorithms \cite{Chekanov:2018nuh,universe7010019}.
Later, it was successfully demonstrated that the same inputs are appropriate for anomaly detection using autoencoders \cite{Chekanov:2021pus,ATLAS:2023ixc,Chekanov:2023uot}. 
Therefore, the RMM inputs are expected to perform equally well for both supervised classifiers and unsupervised autoencoders, which is crucial for comparing these two techniques that adopt the same inputs.

\subsection{Unsupervised autoencoder}

An autoencoder (see \cite{Hinton2006Reducing} and references therein)  is a neural network trained to reconstruct its input while learning a compressed latent representation (latent code) that captures the underlying structure of the data. For anomaly detection, autoencoders are trained on representative normal data, and the reconstruction error (or related likelihood-based measures) is used as an anomaly score, while instances with high scores are flagged as anomalous. 

The autoencoder is implemented using {\sc TensorFlow}~\cite{abadi2016tensorflow}.
It comprises two sections, an encoder and a decoder.
The encoder compresses the input to a latent dimensional space, whereas the decoder takes the data in the latent layer and decompresses it back to its original size. The network contains two hidden layers, with 800 and 400 neurons, respectively, and a latent layer of 200 neurons.
The decoder reverses the structure of the encoder, using 400 and 800 neurons for the two hidden layers, and 1287 neurons for the output layer. 
The Leaky ReLU~\cite{xu2015empirical} activation function is used for all hidden and output layers. 
The autoencoder was trained by minimizing the MSE error using a batch size of 10. With a patience value of 10, the typical number of epochs was 200 for the final training.

The architecture of autoencoder discussed above, as well as the training method,  was found to be the most optimal for performing unsupervised learning \cite{ATLAS:2023ixc,Chekanov:2023uot} using the RMM inputs. The performance can be tested using the ADfilter tool \cite{Chekanov_2025}.

\subsection{Supervised classifier}

In its simplest form, a supervised classifier is a feedforward neural network, typically a multilayer perceptron, trained using backpropagation. It learns to assign inputs to labeled classes. Thus, it requires a well-defined training dataset, such as events annotated as BSM or SM.

For a fair comparison with anomaly detection, the supervised classifier uses the same input as the anomaly detector - namely, the RMM matrix. The use of RMMs as model-agnostic inputs for supervised classifiers has been discussed in Ref. \cite{universe7010019}.

The supervised classifier uses the same RMM input as the unsupervised autoencoder and was also created with {\sc TensorFlow}. The classifier repeats the architecture of the encoding part of the autoencoder, having 2 hidden layers with 800 and 400 neurons each and the ReLU activation function between layers. The output layer has a single neuron that was used to classify the events.

During training, the output layer was set to 0 for SM events and 1 for BSM events.
The same number of SM and BSM events were used, dividing both samples into training and validation sets with a 70/30 ratio. The mass of the scalar $X$ was not included in the inputs, allowing the network to be blind to the physics parameters of the BSM model. Supervised training was performed using only the $X \rightarrow HH$ process, and the $X \rightarrow SH$ process was used to test the model trained on this process.

\section{Results}
\label{sec:results}

\begin{figure}[t]
  \begin{center}
    \subfloat[$X\rightarrow HH$]{\includegraphics[width=0.49\textwidth]{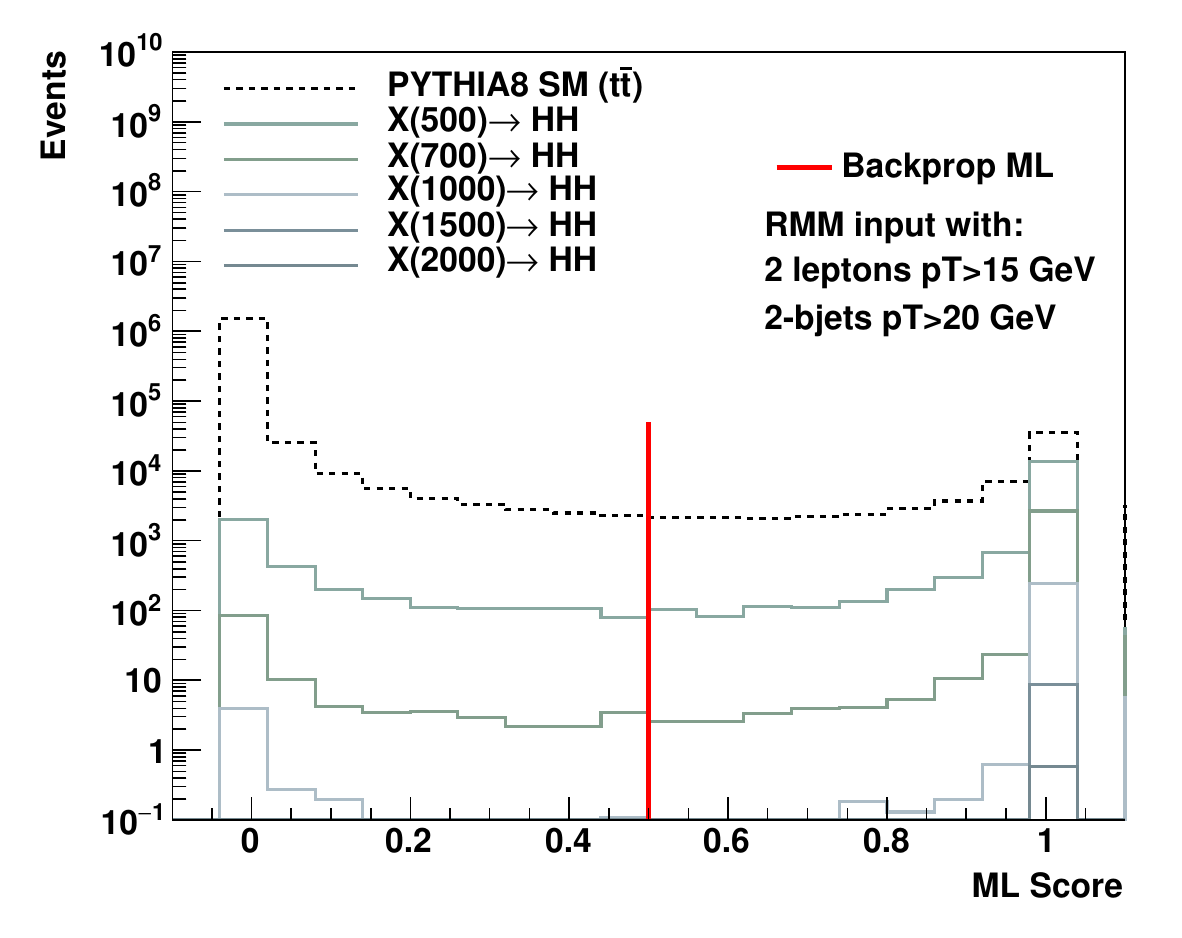}} 
    \subfloat[$X\rightarrow SH$]{\includegraphics[width=0.49\textwidth]{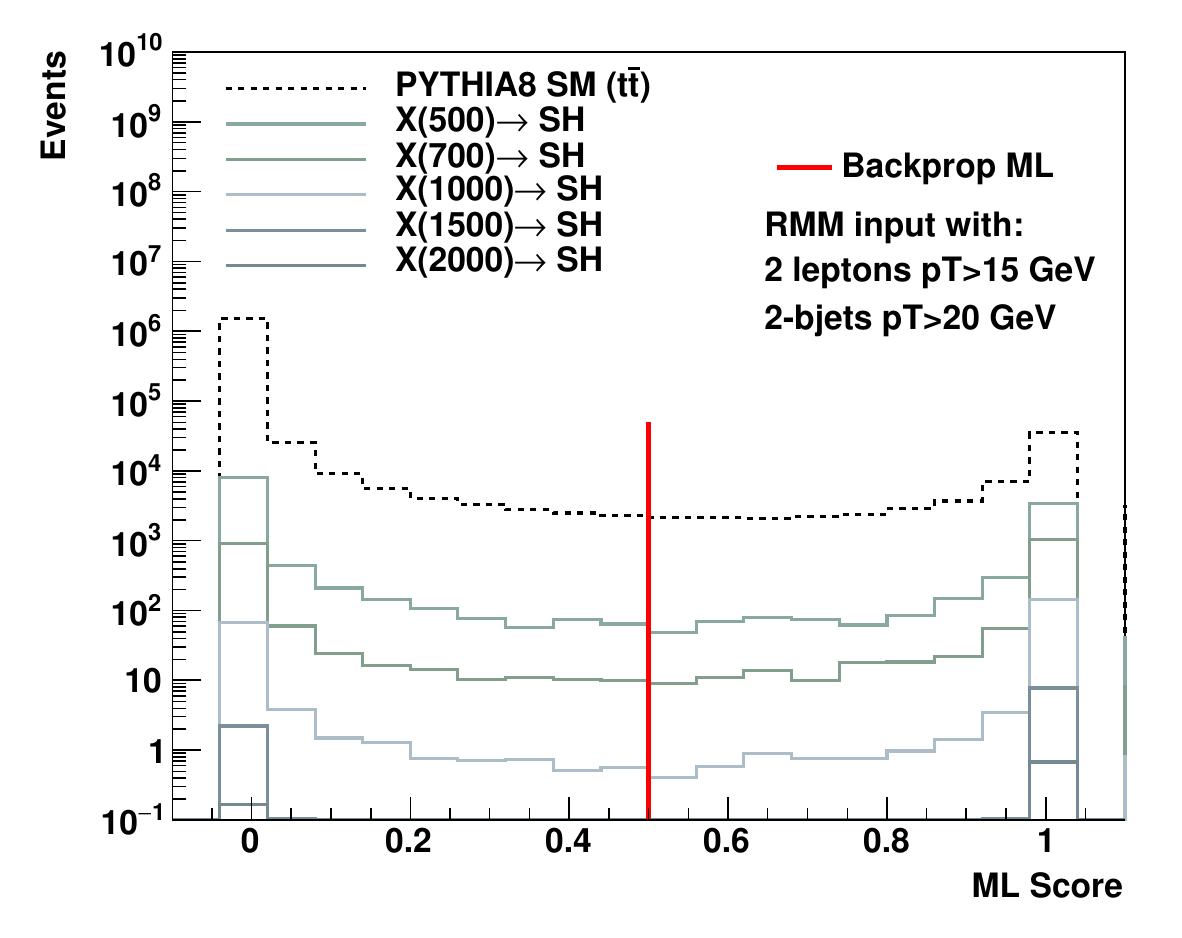}}
  \end{center}
\caption{Scores from the SC for the processes $X\rightarrow HH$ (a)  and  $X\rightarrow SH$ (b). The vertical lines show the threshold values used to select signal events.
}
\label{fig:BPNN_score}
\end{figure}

\begin{figure}[t]  
  \begin{center}
    \subfloat[$X\rightarrow HH$]{\includegraphics[width=0.49\textwidth]{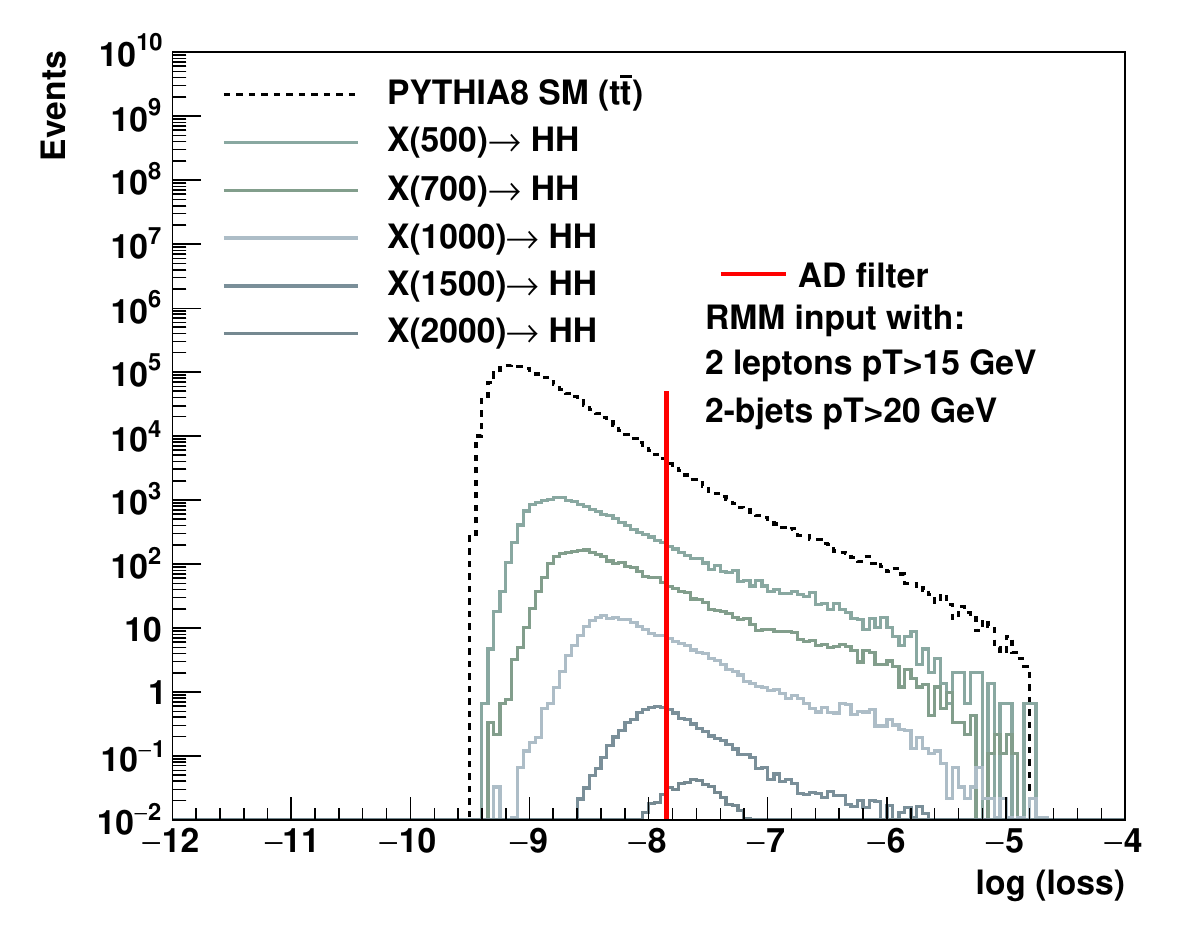}} 
    \subfloat[$X\rightarrow SH$]{\includegraphics[width=0.49\textwidth]{loss_cut_SH.pdf}}
  \end{center}
\caption{Loss values for the anomaly detection used in the processes $X\rightarrow HH$ (a) and  $X\rightarrow SH$ (b).  The vertical red lines show the threshold values used to select anomalous events using the $3\sigma$ rule based on the background events (dashed black lines).
}
\label{fig:loss_cuts}
\end{figure}

Figures~\ref{fig:BPNN_score} and ~\ref{fig:loss_cuts} illustrate the results of a supervised classifier and anomaly detection algorithm applied to the processes \(X \to HH\) and \(X \to SH\), comparing different mass hypotheses for the resonance \(X\) and standard model top-quark production generated with {\sc Pythia8}. 

Figure~\ref{fig:BPNN_score} displays the score distributions for \(X \to HH\) (a) and \(X \to SH\) (b), where the x-axis represents the classifier's confidence in identifying an event as a signal.
The SC score is a scalar discriminant derived from the sigmoid output of a binary neural network classifier. By construction, it ranges from 0 to 1, with values near 0 indicating SM–like (background) events and values near 1 indicating BSM–like (signal) events.
The different colored lines represent various signal masses (500 GeV to 2000 GeV), while the dotted black line corresponds to the Standard Model (SM) background. The red vertical line indicates the classification threshold (0.5), where events to the right are identified as signal-like, demonstrating separation, though variations exist between \(X \to HH\) and \(X \to SH\).

Figure ~\ref{fig:loss_cuts} illustrates the distribution of log values of the reconstruction loss, obtained using the anomaly detection algorithm for \(X \to HH\) (left) and \(X \to SH\) (right). 
The loss values quantify how poorly an event is reconstructed by the network - larger values therefore indicate a higher degree of anomaly.
The same signal mass hypotheses and background distributions are shown, with the red vertical line representing an anomaly detection threshold. These thresholds were determined by calculating the mean values of the loss distributions and by adding three standard deviations to the means. Empirically, this choice represents the  "3$\sigma$ rule", or 0.15\% of anomalous events, assuming the loss distribution follows one-sided normal distribution. Although these thresholds could be tuned to a particular BSM model, we aim for a model-agnostic generalized strategy.

\begin{figure}[t]
  \begin{center}
    \subfloat[\(X \to HH\) after SC]{\includegraphics[width=0.49\textwidth]{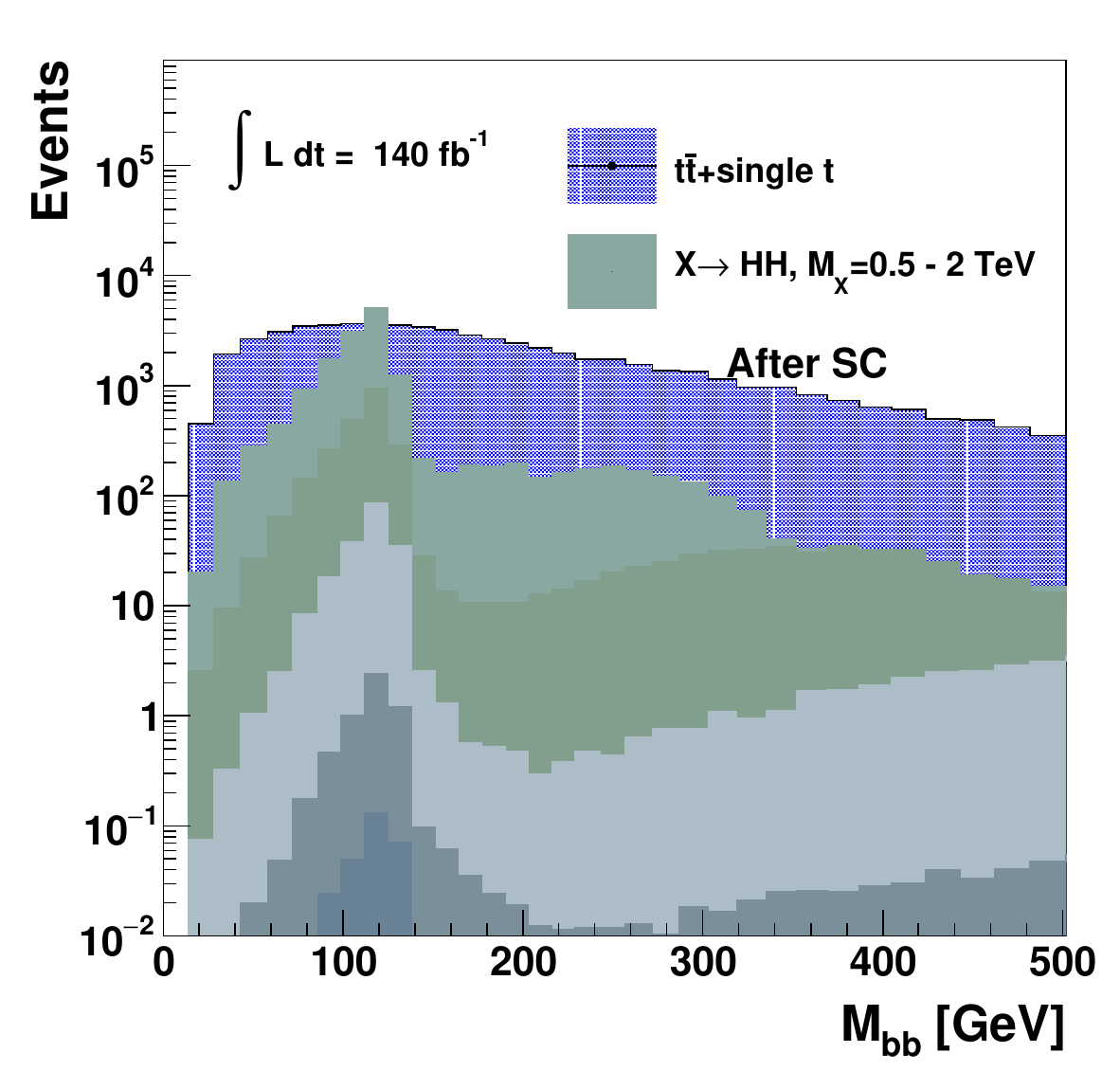}}
 \subfloat[\(X \to HH\) after AD]{\includegraphics[width=0.49\textwidth]{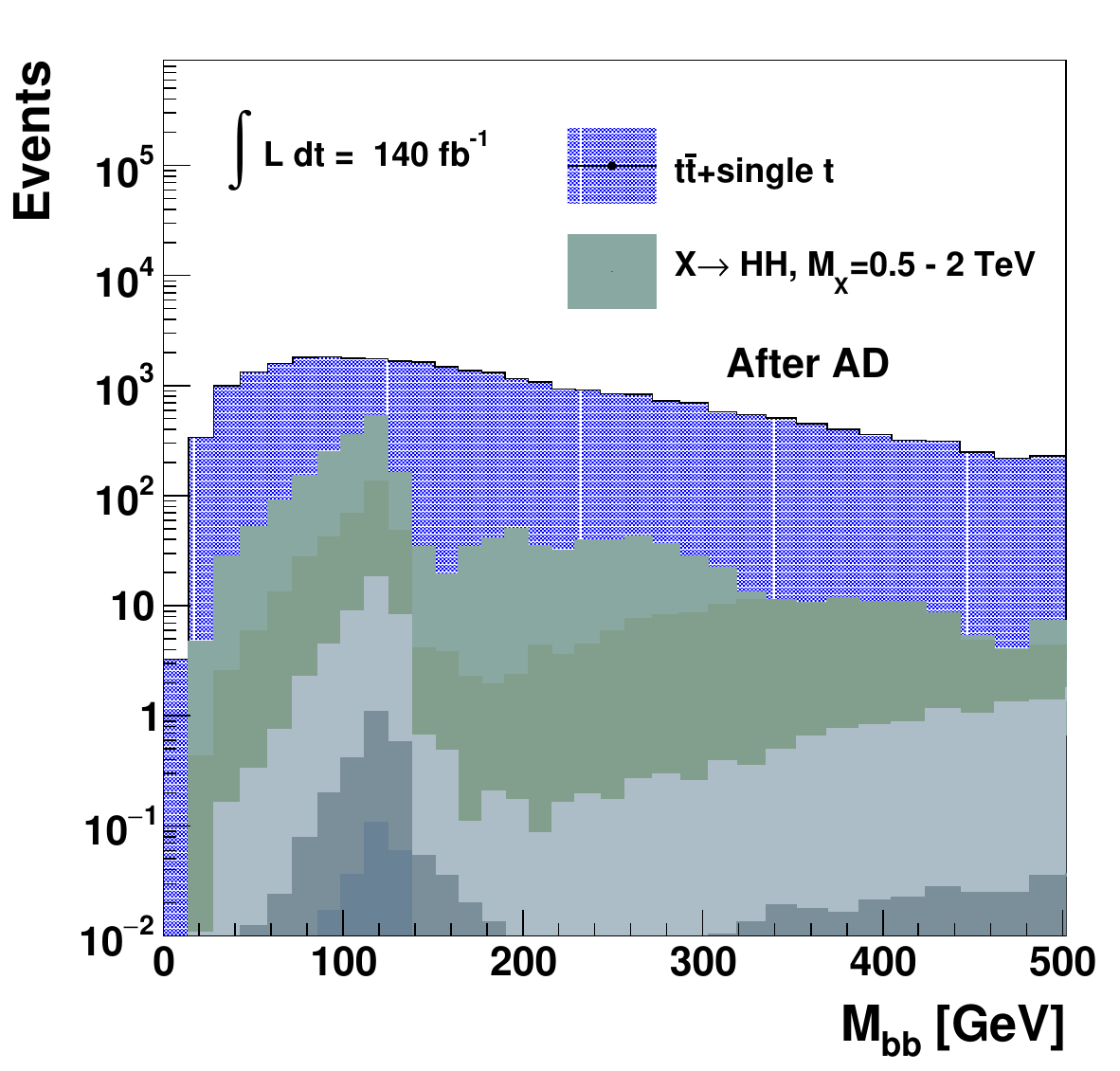}}
  \end{center}
\caption{Distributions for $X\rightarrow HH$ after applying the SC (a) and AD (b) selections.
}
\label{fig:kin_mbb_hh}
\end{figure}

\begin{figure}[H]
  \begin{center}
    \subfloat[\(X \to SH\) after SC]{\includegraphics[width=0.49\textwidth]{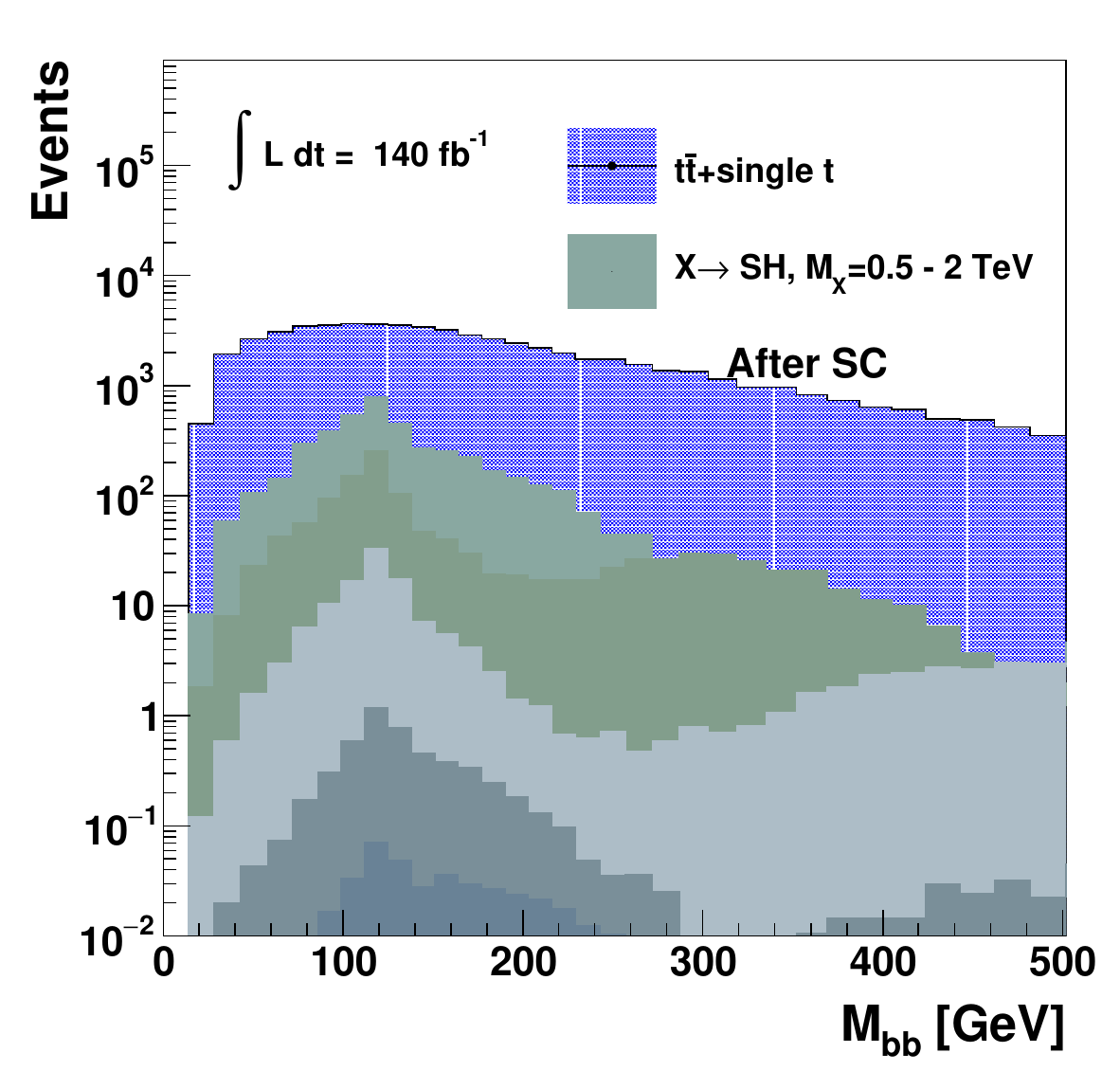}}
 \subfloat[\(X \to SH\) after AD]{\includegraphics[width=0.49\textwidth]{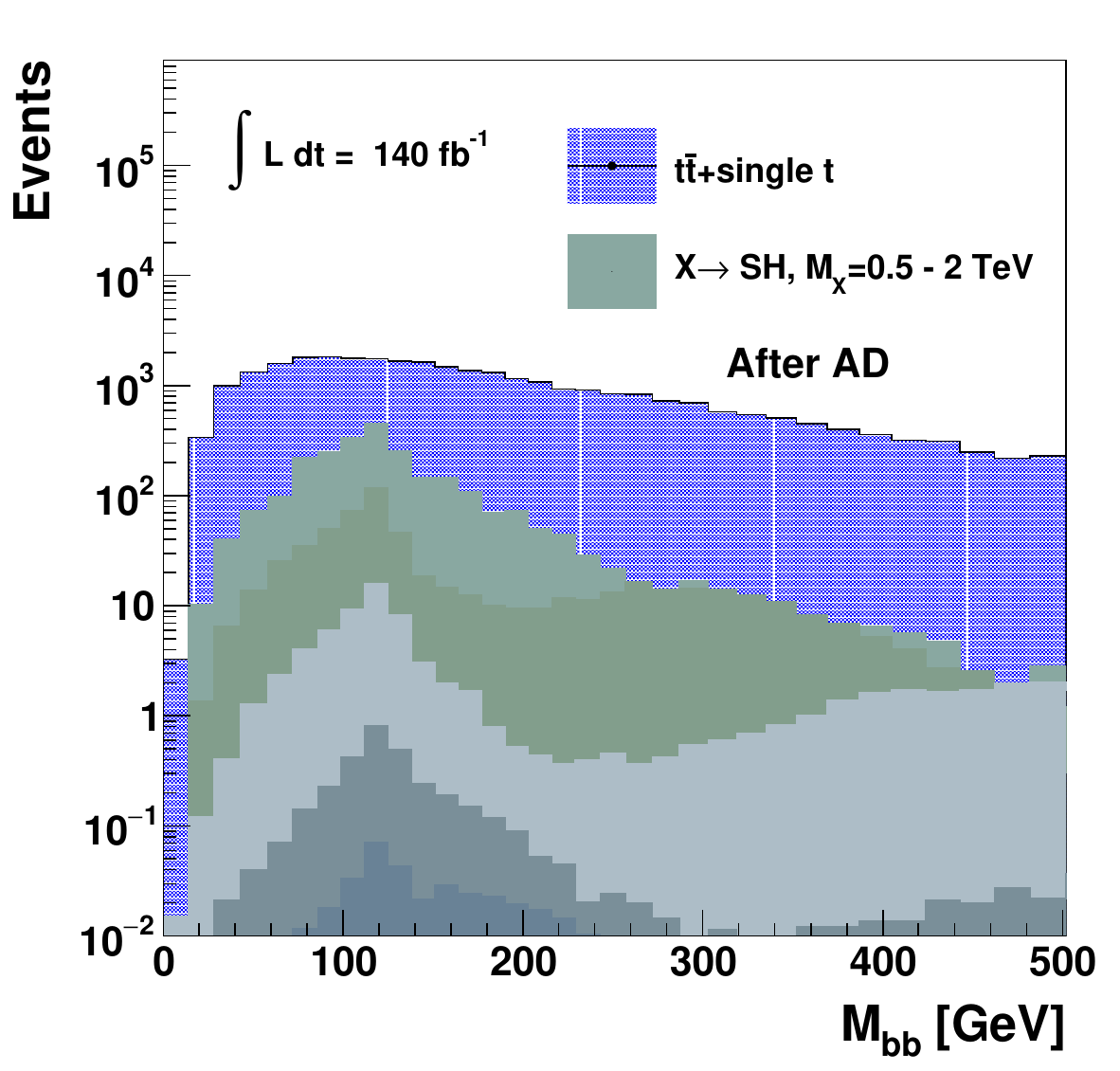}}
  \end{center}
\caption{Distributions for $X\rightarrow SH$, with $S\rightarrow HH$ after applying the SC (a) and AD (b) selections.
}
\label{fig:kin_mbb_Sh}
\end{figure}

Figures ~\ref{fig:kin_mbb_hh} and ~\ref{fig:kin_mbb_Sh} show the distributions for \(X \to HH\) and \(X \to SH\) after applying ML using the SC and AD, respectively. In the \(X \to HH\) case, the SC accepts a slightly larger number of backgrounds events than the AD. For smaller $X$ mass points, the SC accepts a significantly larger number of signal events than the AD. The difference in signal acceptance between the models gradually decreases until the highest mass point 2 TeV, where they accept an almost identical number of events. The \(X \to SH\) case is largely similar, though the difference between the SC and AD is smaller for both signal and background.

In general, we do not expect the SC method to outperform the AD method for processes such as
\(X \to SH\) as seen in Fig.~\ref{fig:kin_mbb_Sh}, 
since this channel was not included in the SC training.
Nevertheless, the performance of the SC method for this process is remarkably similar to that of the AD approach.
This suggests that the classifier is able to capture the key kinematic features of the \(X \to SH\). 
This could be due to similarity of the hadronic final state in the \(X \to HH\) and \(X \to SH\) processes.

To better quantify the performance differences, we evaluated the significance as a function of the particle mass of the particle $X$. The significance $Z$ is calculated \cite{Cowan:2010js} as 

$$
Z=\sqrt{2 \left(  (S+B)\ln \left( 1+\frac{S}{B} \right) -S \right)},
$$
where $S$ is the number of signal events and $B$ is the number of background events. 
Both numbers were calculated in the region of the $\mbb$ distribution near the nominal Higgs boson mass, using a 50 GeV invariant-mass window that contains the majority of the signal entries. 
This mass window was not optimized, since the analysis is limited to a comparison
between  $\mbb$ invariant masses after  SC and AD. 

\begin{figure}[H]
  \begin{center}
    \subfloat[$X \to HH$ ]{\includegraphics[width=0.49\textwidth]{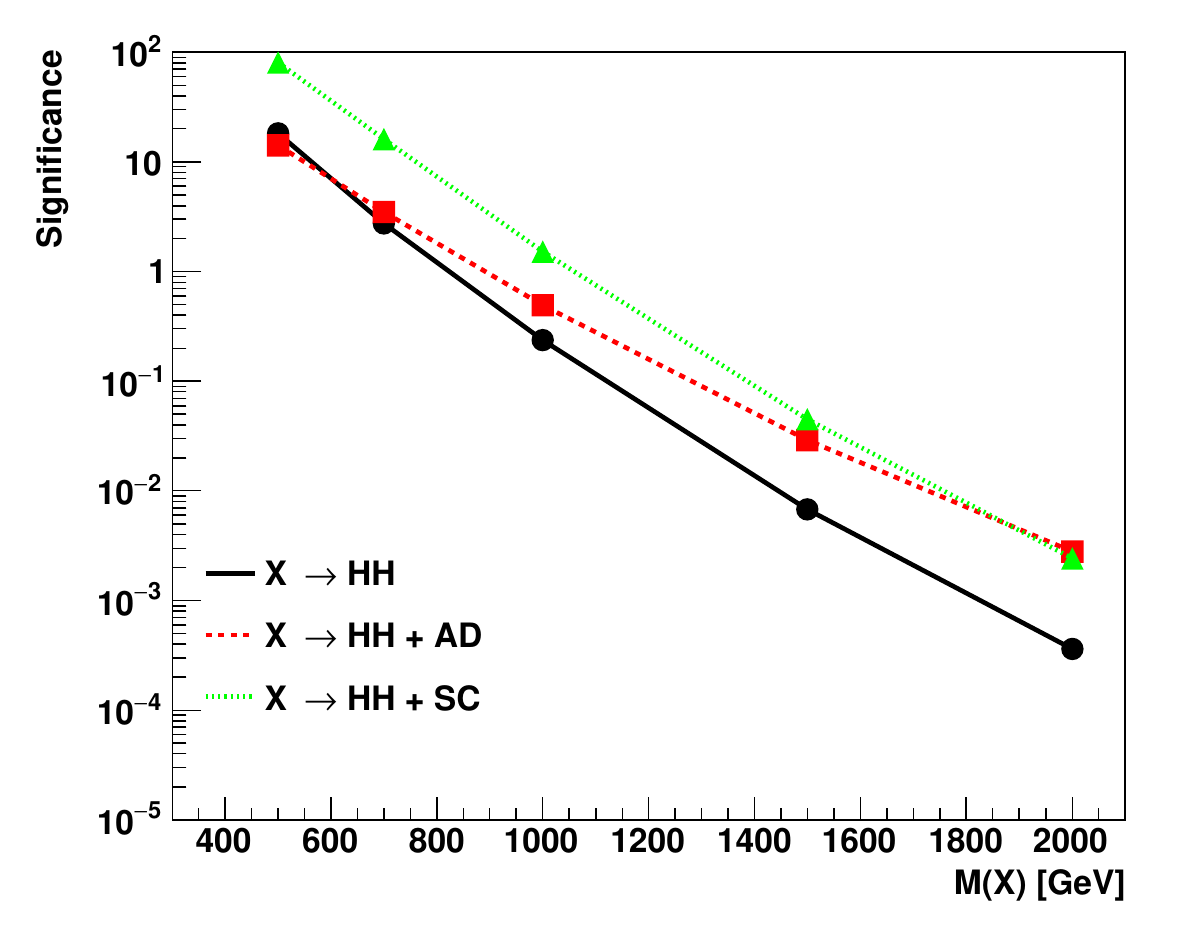}} 
    \subfloat[$X \to SH$]{\includegraphics[width=0.49\textwidth]{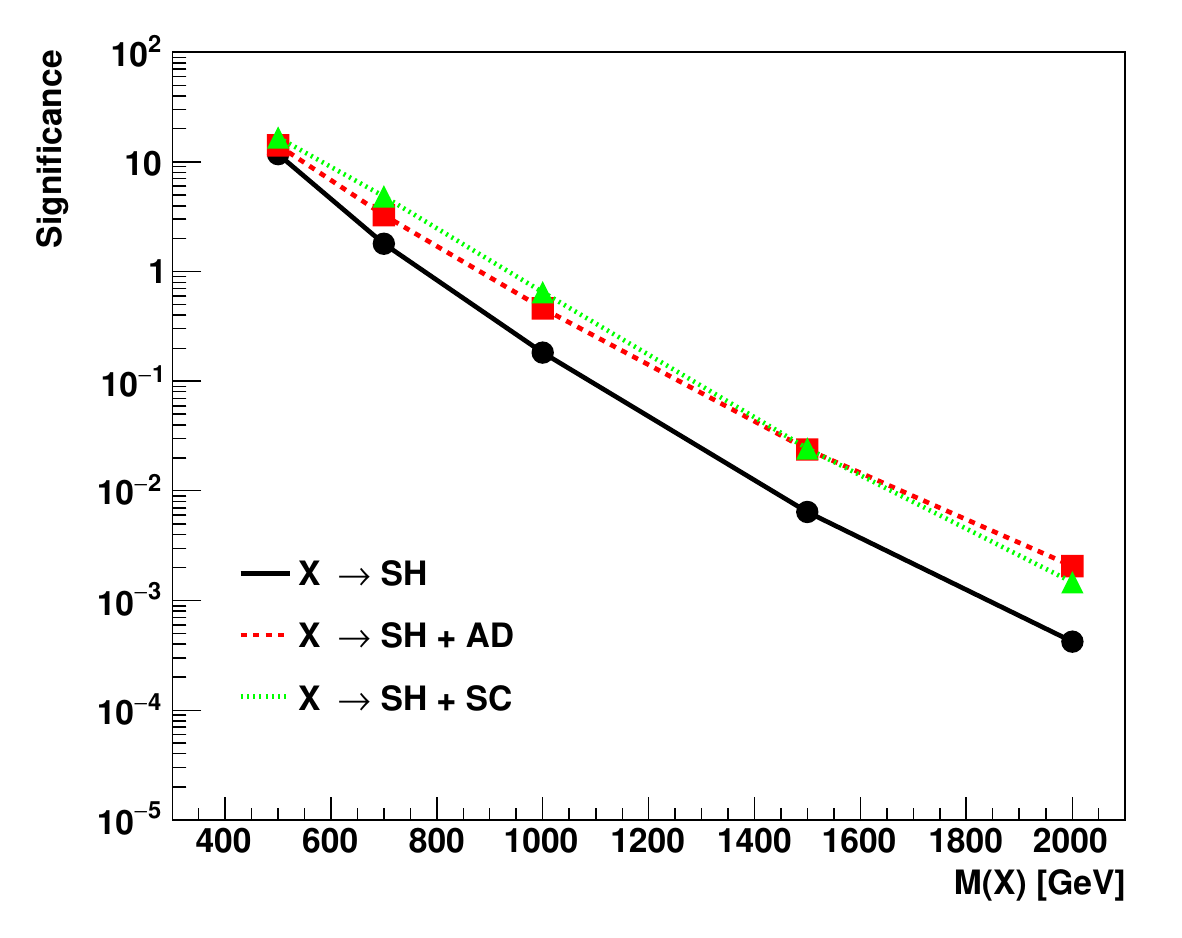}}
  \end{center}
\caption{Comparison of the significance values for
(a) $X\rightarrow HH$; (b) $X\rightarrow SH$, with $S\rightarrow HH$ channels. See the text for the definition of the significance. 
}
\label{fig:signi_cxSM}
\end{figure}

Figure~\ref{fig:signi_cxSM} shows that the significance is enhanced using the SC and AD for both \(X \to HH\) and \(X \to SH\) channels. For the \(X \to HH\) channel, the supervised event classifier is more effective at lower masses and for higher masses significance is enhanced using AD as well.  
For the \(X \to SH\) channel, at higher masses, AD outperforms the SC. 
It is expected that two b-jets from a single Higgs boson are boosted at large masses of $X$. This may reduce the efficiency of reconstruction of $\mbb$, but this effect should be common for both SC and AD reconstructions, and thus should not affect our conclusion about the comparison of these two ML approaches.

\section{Conclusion}
\label{sec:conclusion}

In summary, we tested two fundamentally different  ML approaches: one, a supervised classifier that requires both SM background and BSM signal events as input, and the other, an autoencoder-based anomaly detection approach that only requires SM events for training. The benchmarks were based on models with $X\rightarrow HH$ and $X\rightarrow SH$ boson decays, with the final state involving two leptons and two $b$-jets. The same input variables were used in both ML approaches, and the architecture of the hidden layers was identical. Therefore, these studies provide a compelling comparison of two distinct approaches that use the same inputs.  Such a comparison of very different methods, using the same input variables, has been performed for the first time. Note that the anomaly-detection technique has not been used so far in di-Higgs searches. The Monte Carlo data and the analysis code for this paper are available from \cite{Chekanov:2014fga} and \cite{github}. 

We conclude that both ML techniques were able to improve the significance of signal observations. The event classifier outperforms the autoencoder for the BSM events used in training, especially for low masses of the scalar $X$. This is unsurprising for the $HH$ signals, which were used to train the SC. However, it was unexpected to observe that the classifier works quite well even for the $SH$ signal events, which were not used in training. Its performance is comparable to that of the autoencoder. 
The latter begins to slightly outperform the supervised classifier at higher values of the $X$ mass. The good agreement between the SC and AD in the $SH$ channel is likely due to the relatively similar final states predicted by these BSM models.

It is expected that the autoencoder starts to outperform the SC for events that have a substantial difference in the final state, such as the presence of additional identified particles that are not part of the Higgs decays.  Thus, the method offers a model-agnostic approach that generalizes more effectively across multiple BSM scenarios compared to a supervised classifier in di-Higgs analyses targeting high-mass resonances.



\section{Acknowledgments}
We would like to thank R.~Zhang, D.~Chen, I.~Low for the discussion of the anomaly detection  technique 
and for providing the Madgraph model cxSM\_VLF\_EFT for the $HHH$ signals.

The submitted manuscript has been created by UChicago Argonne, LLC, Operator of Argonne National Laboratory (“Argonne”). Argonne, a U.S. 
Department of Energy Office of Science laboratory, is operated under Contract No. DE-AC02-06CH11357. 
Argonne National Laboratory’s work was 
funded by the U.S. Department of Energy, Office of High Energy Physics (DOE OHEP) under contract DE-AC02-06CH11357.  We gratefully acknowledge the computing resources provided by
the Laboratory Computing Resource Center at Argonne National Laboratory.

\newpage

\bibliographystyle{JHEP}
\bibliography{references}






\end{document}